\documentclass[
reprint,
superscriptaddress,
amsmath,amssymb,
aps, physrev,
]{revtex4-2}

\usepackage{graphicx}
\usepackage{dcolumn}
\usepackage{bm}
\usepackage{braket}
\usepackage{siunitx}
\usepackage{comment}

\begin{document}

\title{\textbf{Emulating isomerization with two-dimensional Coulomb crystals}}

\author{Naoto Mizukami}
\email{Contact author: mizukami@lens.unifi.it}
\affiliation{Istituto Nazionale di Ricerca Metrologica, Strada delle Cacce 91, 10135 Torino, Italy}
\affiliation{European Laboratory for Nonlinear Spectroscopy (LENS), Via Nello Carrara 1, 50019 Sesto Fiorentino}
\affiliation{Politecnico di Torino, 10129 Torino, Italy}
\author{Gabriele Gatta}
\affiliation{European Laboratory for Nonlinear Spectroscopy (LENS), Via Nello Carrara 1, 50019 Sesto Fiorentino}
\affiliation{University of Florence, Department of Physics and Astronomy, Via Sansone 1, 50019 Sesto Fiorentino, Italy}
\author{Lucia Duca}
\affiliation{Istituto Nazionale di Ricerca Metrologica, Strada delle Cacce 91, 10135 Torino, Italy}
\affiliation{European Laboratory for Nonlinear Spectroscopy (LENS), Via Nello Carrara 1, 50019 Sesto Fiorentino}
\author{Carlo Sias}
\email{Contact author: c.sias@inrim.it}
\affiliation{Istituto Nazionale di Ricerca Metrologica, Strada delle Cacce 91, 10135 Torino, Italy}
\affiliation{European Laboratory for Nonlinear Spectroscopy (LENS), Via Nello Carrara 1, 50019 Sesto Fiorentino}
\affiliation{Istituto Nazionale di Ottica, Consiglio Nazionale delle Ricerche (CNR-INO), Largo Enrico Fermi 6, 50125 Firenze, Italy}

\date{\today}

\begin{abstract}
Isomerization, i.e. the rearrangement between distinct molecular configurations, is a fundamental process in chemistry. Here we demonstrate that two-dimensional Coulomb crystals can emulate molecular isomerization and be used to characterize its physical mechanisms. In our molecular analogue, the confining potential acts as an electronic orbital, which can be tuned continuously and dynamically. We use a planar crystal of six $^{138}$Ba$^+$ ions, which exhibits two stable configurations depending on the aspect ratio of the harmonic trapping potential. By changing this aspect ratio, we directly modify the potential energy surface (PES) of the ion crystal, and trigger isomerization in a controlled way. We identify a region of bistability between the two isomers, and use configuration-resolved imaging to detect isomerization in real time. A Monte Carlo simulation is used to calculate the double-well PES. By comparing simulated transition rates with experimental population ratios, we estimate the crystal’s temperature. Additionally, we prepare metastable configurations by rapidly quenching the PES, and detect isomerization dynamics with sub-millisecond resolution. Our work establishes a new platform for emulating molecular processes, paving the way for studying quantum superpositions of crystal configurations, and for controlling isomeric excitations in two-dimensional Coulomb crystals.
\end{abstract}

\maketitle

Atomic systems have been used successfully over the past few decades for simulating a wide range of physical phenomena. This progress is largely due to the exceptional degree of control over key parameters that govern the particles' dynamics, such as their confinement and interactions. Both neutral atoms and trapped ions have been used to emulate physical models e.g. of condensed matter \cite{Bloch2012, Blatt2012}, high energy physics \cite{PRXQuantum.4.027001}, and astrophysics \cite{PhysRevLett.131.223401, PhysRevD.107.023007}, by replicating the corresponding Hamiltonians. This success stems not only from the precise control of particle dynamics, but also from the fact that the characteristic timescales in atomic systems can be made much more accessible than for other physical systems, e.g. electrons in a solid. For instance, Bloch oscillations, a phenomena occurring at a picosecond timescale for electrons in a superlattice \cite{Feldmann1992}, lies in the millisecond range for atoms in an optical lattice \cite{Morsch2001}.

Trapped ions, when cooled to sufficiently low temperatures, form Coulomb crystals: ordered structures shaped by mutual repulsion and external confinement \cite{PhysRevLett.59.2935, DREWSEN2015105}. These structures resemble molecular configurations, where the role of electronic orbitals is effectively played by the trapping potential \cite{wineland1987}. This analogy is particularly striking in two-dimensional crystals. For example, five or six ions in a symmetric two-dimensional potential can arrange themselves in a geometry that is reminiscent of an aromatic ring in a molecule, like in pyrrole or benzene \cite{duca2023}. Similarly to a molecule, a two-dimensional crystal of ions has an energy spectrum characterized by a vibrational and a rotational quantum number \cite{Urban2019}. However, the size of an ion crystal is approximately five orders of magnitude larger than that of a benzene molecule, resulting in an energy spectrum that is much shallower. As a result, the internal dynamics of the crystal occurs on far more accessible timescales with respect to the one of a molecule for its real-time observation.

Like molecules, two-dimensional ion crystals can exhibit isomerization, i.e. changes in the spatial arrangement of the particles that are triggered by variations in the confining potential. A notable example is the linear to zigzag transition \cite{fishman2008}, which was investigated in different contexts, e.g. to study its second-order phase transitions nature \cite{kiethe2021}, and to investigate the formation of topological defects \cite{Ulm2013}.
Moreover, there can be conditions in which multiple configurations are stable. An example is a quasi-2D crystal, in which the particles can be either in a zigzag or a zagzig configuration, both having the same energy. This physical system was suggested for creating quantum superposition between crystal configurations \cite{retzker2008}. 

In real molecules, isomerization occurs when the molecule re-arrange on a different minimum of the potential energy surface (PES), with a process that can be triggered in different ways, e.g. via thermal excitation or quantum tunneling \cite{Choudhury2022}. The PES can be modified by multiple mechanisms, e.g. via laser-driven population of electronically excited states \cite{vogt2005,Kuebel2020}, the use of mechanical stress \cite{Stauch2016}, the application of electric fields \cite{Aragonès2016}, or the use of different solvents \cite{Widmer2018}. In Coulomb crystals, isomerization can be triggered by changing the external confinement \cite{Yan2016}. In particular, the number of different stable isomers increases with the number of particles \cite{PhysRevLett.71.2753, Mitchell1998}, and the presence of several stable crystal configurations were observed for a number of ions approaching $100$ \cite{Kiesenhofer2023}.

In our work, we use a two-dimensional Coulomb crystal to emulate the process of isomerization in a molecule. Isomerization is triggered by a change of the PES, which is realized by controlling the aspect ratio of the 2D harmonic trapping potential. To do so, we tune the DC voltage applied to the endcaps of our Paul trap \cite{Perego2020}. As this parameter changes, the particles remain confined in a plane and the trap aspect ratio $\alpha = \omega_y/\omega_z$ changes \cite{duca2023}, where $\omega_z$ and $\omega_y$ are the trap frequencies along the two axis of the confinement plane.
As a result, it is possible to quench the aspect ratio, and observe in real time the dynamics of the crystal while undergoing isomerization.
This property results particularly interesting to investigate bistability regions, in which more than one isomer is populated, with the particles moving between one isomer to the other triggered by thermal fluctuations. 

In our study, we focus on the specific case of six $^{138}$Ba$^+$ ions forming a two-dimensional Coulomb crystal. We identify a range of parameters in which the crystal undergoes isomerization by passing from a structure formed by six ions in a benzene-like ring, to a crystal formed by five ions in a ring and one ion at the center of the trap. We will refer to these configurations as ``hexagon'' and ``pentagon'' isomers, respectively. 
This specific example has the advantage that isomerization is controlled by the displacement of one of the ions in the crystal from the center of the trap to the outer ring. As a result, a single parameter can be used to characterize the PES, which takes the form of an energy double well, in which we can precisely control the energy mismatch between the wells.  
We quench the PES by quickly ramping the electric potential of the trap, and identify a bistability region. We characterize this region, and extract the temperature of the crystal. Moreover, we observe in real-time the occurrence of isomerization, and extract its timescale as a function of the trap aspect ratio.
This data is used to extract the amplitude of the transition rates between the two isomers, thus fully characterizing the isomerization process. 
In parallel with the experiment, we perform a Monte Carlo simulation \cite{kong2002, PhysRevB.66.155322}, and compare the results with the experimental data, finding an excellent agreement.

\subsection*{Locating bistable region with a Monte Carlo simulation}
\begin{figure}[t]
\includegraphics[width=\linewidth,pagebox=cropbox]{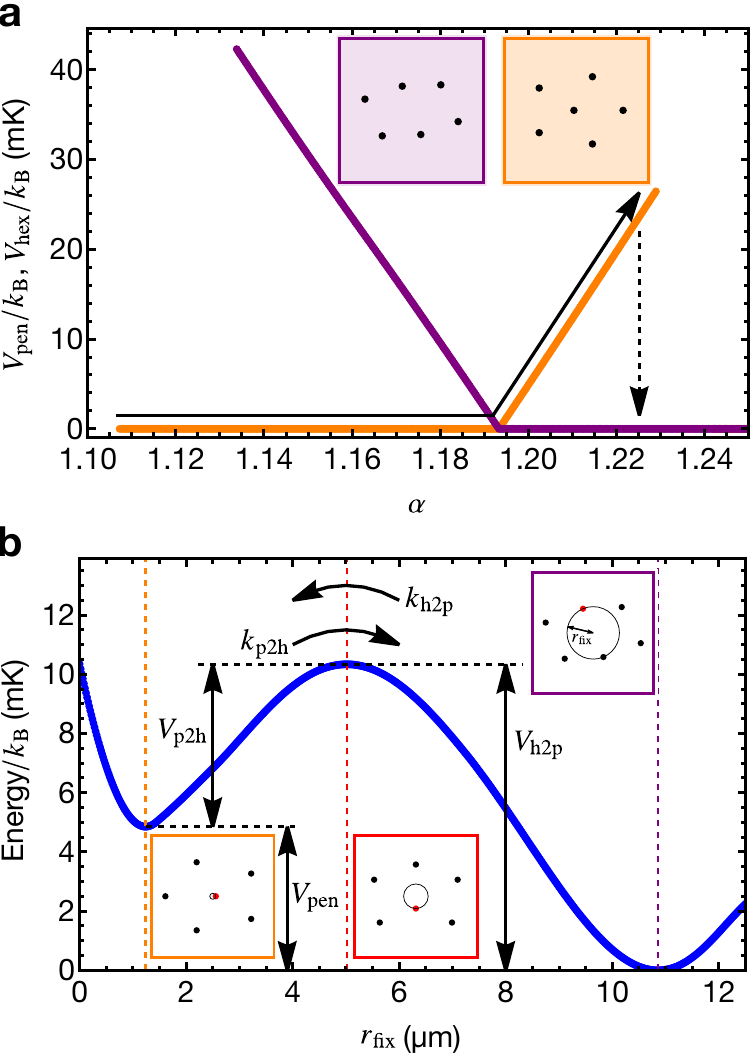}
\caption{\label{fig:bistability}
{\bf{Simulation of two isomers of six ions.}}
{\bf a} Potential energies of the pentagon (orange) and hexagon (purple) isomers as a function of the trap aspect ratio $\alpha$.
By quenching $\alpha$, it is possible to populate a metastable isomer (black arrow), which then decays by thermal excitations to the lowest energy one (dashed arrow).
{\bf b} Energy along the minimum energy path on the potential energy surface for $\alpha = 1.20$. The minimum energy is extracted by forcing one ion to stay at a distance $r_{\mathrm{fix}}$ from the trap center (see insets) in the Monte Carlo simulation. The calculated energies for the two minima and the saddle point make it possible to extract the barrier heights $V_{\mathrm{p2h}}$, $V_{\mathrm{h2p}}$ and the transition rates $k_{\mathrm{p2h}}$, $k_{\mathrm{h2p}}$.
In the plots, an energy offset is subtracted to set the lowest energy to zero.
}
\end{figure}

In order to identify the bistability region and extract the corresponding PES, we use a Monte Carlo simulation.
In the simulation (see methods), the equilibrium position of six ions in the trap is found by an iterative process. In each step of this process, two random particles are moved around their initial position and displaced to the position corresponding to the lowest energy. 
In an initial set of simulations, we extracted the ions’ equilibrium positions and their total potential energy (confinement and electrostatic energies) as a function of the aspect ratio $\alpha$. This was done by gradually increasing $\alpha$, using the equilibrium configuration from the previous step as the starting point for each new simulation. As $\alpha$ increases, isomerization occurs at a threshold value, corresponding to an abrupt switching of the ions' equilibrium position from the pentagon to the hexagon isomer. After reaching the upper limit of $\alpha$, we reversed the process and gradually decreased $\alpha$. Notably, the results revealed pronounced hysteresis, with the isomerization threshold and equilibrium energies depending on the direction of the sweep. This behavior signals the presence of bistability between the two distinct isomers.
Fig. \ref{fig:bistability}a summarizes the results of this simulation by plotting the potential energies $V_{\mathrm{pen}}$ and $V_{\mathrm{hex}}$ at equilibrium for the pentagon and hexagon isomer, respectively. 
For $\alpha \lesssim 1.19$ the lowest energy configuration corresponds to the pentagon isomer (orange), while for $\alpha \gtrsim 1.19$, the lowest energy configuration is the hexagon isomer (purple).
Interestingly, for $1.13\lesssim \alpha \lesssim 1.23$ both isomers can be prepared as a stable solution of the Monte Carlo simulation, a clear signature of bistability.
Crucially, the isomerization scale lies in the $10$s \si{\milli\kelvin} level, i.e. values that can be explored with typical trapped ion kinetic energies.

The difference between the two isomers can be characterized by the minimum distance $r_{\mathrm{min}}$ between an ion and the trap center.
During the isomerization process, $r_{\mathrm{min}}$ changes abruptly as one ion displaces from or to the center of the confining potential.
To extract the PES along the isomerization path, we ran a second set of Monte Carlo simulations.
In this set, we forced one ion to stay at a specific distance $r_{\mathrm{fix}}$ from the center of the trap.
As $r_{\mathrm{fix}}$ is varied, the PES is built up from each crystal configuration corresponding to the energy minimum.
Fig. \ref{fig:bistability}b plots the energy of six ions obtained with the Monte Carlo simulation as a function of $r_{\mathrm{fix}}$ for $\alpha=1.20$, corresponding to $(\omega_z, \omega_y)/(2\pi) = (100, 120)$ \si{\kilo \hertz}.
The energy along the minimum energy path shows a double well behaviour, typical to the PES in the presence of bistability.
We extract the value of the energy barrier between the pentagon (hexagon) and the hexagon (pentagon) isomers $V_{\mathrm{p2h}}$ ($V_{\mathrm{h2p}}$) as the energy difference between the minimum energy of the isomers and the energy of the saddle point $V_{\mathrm{sad}}$ thus given as $V_{\mathrm{p2h}} = V_{\mathrm{sad}} - V_{\mathrm{pen}}$ and $V_{\mathrm{h2p}} = V_{\mathrm{sad}} - V_{\mathrm{hex}}$.
We observe the presence of a double well in the PES in the similar region of $\alpha$ in which we observe hysteresis in the first set of simulations.

\subsection*{Formation and detection of a metastable isomer}
Experimentally, we exploit the hysteresis shown in the Monte Carlo simulation to configure ions into a metastable isomer.
\begin{figure}[b]
\centering
\includegraphics[width=\linewidth,pagebox=cropbox]{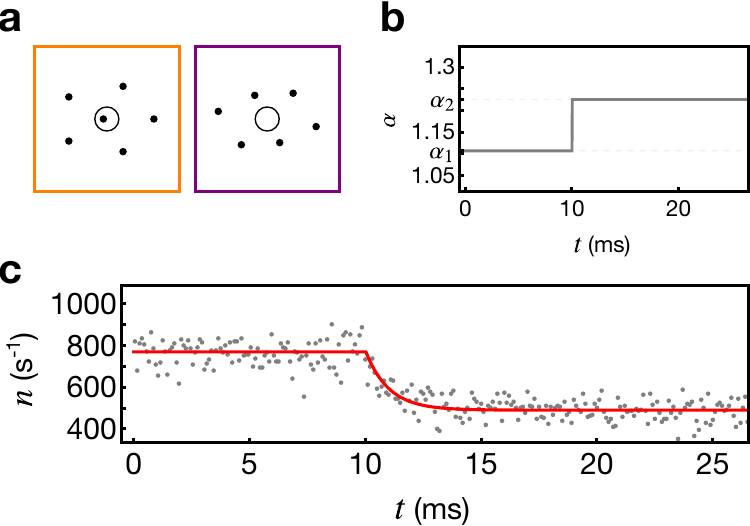}
\caption{\label{fig:decay}
{\bf{Real-time observation of isomerization.}}
{\bf{a}} Monte Carlo generated equilibrium position of the ions in the pentagon and hexagon isomers. A configuration-resolved imaging is set by observing the fluorescence only from the center of the trap (black circle). 
{\bf{b}}
Typical experimental run. The trap aspect ratio $\alpha$ is quenched  from $\alpha_{1} = 1.107$ to the variable value $\alpha_2$ at $t=$10 \si{\milli \second}.
{\bf{c}} Photon rate of the fluorescence observed in 2072 sequences for $\alpha_{2} = 1.225$. The fluorescence detected from the central ion (circle in {\bf{a}}) decreases exponentially after quenching ($t>10$ms), witnessing the relaxation from the initial pentagon isomer to a bistable regime.
}
\end{figure}
Fig. \ref{fig:decay}b shows a typical experimental run.
The ions are first prepared in a pentagon isomer by setting $\alpha$ to a value $\alpha_1 = 1.11$ at which no metastable configuration is present.
After $10$ \si{\milli \second} of laser cooling, $\alpha$ is rapidly quenched in less than $50$ \si{\micro\second} to a value $\alpha_{2}$.
For $1.13 \lesssim \alpha_2 \lesssim 1.23$, the crystal experiences bistability between the hexagon and the pentagon isomers. 
The crystal is kept at $\alpha = \alpha_2$
for $20$ \si{\milli \second} before it is changed back to $\alpha_1$ and a new cycle can begin. Laser cooling remains on for the whole cycle, and we can assume the ions to be at the same temperature in their steady state isomer. 
To distinguish between the two configurations, we have implemented a setup to detect fluorescence from only the center region of the trap (see Fig. \ref{fig:decay}a and methods).
We identify what isomer is formed by the ions by measuring the photon counting rate $n$.
Since the pentagon isomer is characterized by one ion in the trap central region, the corresponding photon rate $n_{\mathrm{pen}}$ is larger than the photon rate $n_{\mathrm{hex}}$ measured in the case of a hexagon isomer.
Whenever bistability occurs, the crystal continuously jumps from one isomer to the other because of thermal motion, and we observe a photon rate $n_{\mathrm{bis}}$ that lies between the two limiting cases, i.e. $n_{\mathrm{hex}} \leq n_{\mathrm{bis}}\leq n_{\mathrm{pen}}$.
Fig. \ref{fig:decay}c shows the average photon rate detected for multiple experimental runs.
After the quench in the aspect ratio we observe an exponential decay of the photon rate to $n_{\mathrm{bis}}$, a signature of the occurrence of isomerization. 

In the bistability region, the crystal continuously hops from one isomer state to the other as an effect of thermal fluctuations. At each given time, it is possible to define a probability $p_{\mathrm{hex}}$ ($p_{\mathrm{pen}}$) for the crystal to be in the hexagon (pentagon) isomer.
These probabilities can be estimated with the photon rate, since its value integrated over time is the result of contributions from the time the crystals spends in each isomer, $n_{\mathrm{bis}}=p_{\mathrm{hex}}n_{\mathrm{hex}}+p_{\mathrm{pen}}n_{\mathrm{pen}}$.
As a result, the probabilities associated with each isomer read:
\begin{equation}
\begin{aligned}
p_{\mathrm{hex}}=\frac{n_{\mathrm{pen}}-n_{\mathrm{bis}}}{n_{\mathrm{pen}}-n_{\mathrm{hex}}},
\\
p_{\mathrm{pen}}=\frac{n_{\mathrm{bis}}-n_{\mathrm{hex}}}{n_{\mathrm{pen}}-n_{\mathrm{hex}}}.
\end{aligned}
\label{Eq:probvsrate}
\end{equation}

\begin{figure}[t]
    \centering
    \includegraphics[width=0.95\columnwidth]{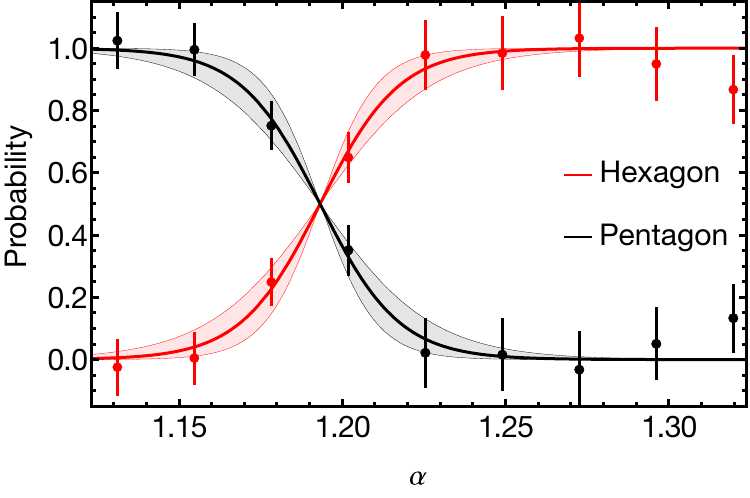}
    \caption{\textbf{Bistability of crystal isomers.} Probability for a crystal to be arranged in a pentagon (red points) or hexagon (black points) isomer as a function of the trap aspect ratio. The lines are obtained from the simulation data, with the temperature being the only free parameter. The shaded area enclose the functions for which $\chi^2$ is twice as big as the optimal one. The corresponding temperature estimate range is $8.8 \substack{+3.5 \\ -2.6}$ \si{\milli\kelvin}. 
   }
    \label{fig_probability}
\end{figure}

Fig. \ref{fig_probability} shows the probabilities $p_{\mathrm{hex}}$ and $p_{\mathrm{pen}}$ as a function of $\alpha$. 
In the bistable regime, the probabilities $p_{\mathrm{hex}}$, $p_{\mathrm{pen}}$ are related to the transition rates by the relation \cite{RevModPhys.62.251}:
\begin{equation}
    \frac{p_{\mathrm{pen}}}{p_{\mathrm{hex}}}=\frac{k_{\mathrm{h2p}}}{k_{\mathrm{p2h}}},
    \label{eq:prob}
\end{equation}
where $k_{\mathrm{h2p}}$ ($k_{\mathrm{p2h}}$) is the transition rate for the crystal changing from the hexagon (pentagon) to the pentagon (hexagon) isomer (see Fig. \ref{fig:bistability}b).
Following Arrhenius equation, the transition rates are related to the temperature $T$ via the relation $k_{\mathrm{h2p(p2h)}}\propto \mathrm{exp}(-V_{\mathrm{h2p(p2h)}}/k_\mathrm{B} T)$, where $k_\mathrm{B}$ is the Boltzmann constant.
Therefore, we can use Eq. (\ref{eq:prob}) to derive the probabilities $p_{\mathrm{hex}}$, $p_{\mathrm{pen}}$ as a function of the barrier height and the temperature:
\begin{equation}
\begin{aligned}
p_{\mathrm{hex}}=\frac{1}{1+\exp[(V_{\mathrm{p2h}}-V_{\mathrm{h2p}})/(k_\mathrm{B} T)]},
\\
p_{\mathrm{pen}}=\frac{1}{1+\exp[(V_{\mathrm{h2p}}-V_{\mathrm{p2h}})/(k_\mathrm{B} T)]}.
\end{aligned}
\label{Eq:probfitting}
\end{equation}

\begin{figure}[b]
    \centering
    \includegraphics[width=0.95\columnwidth]{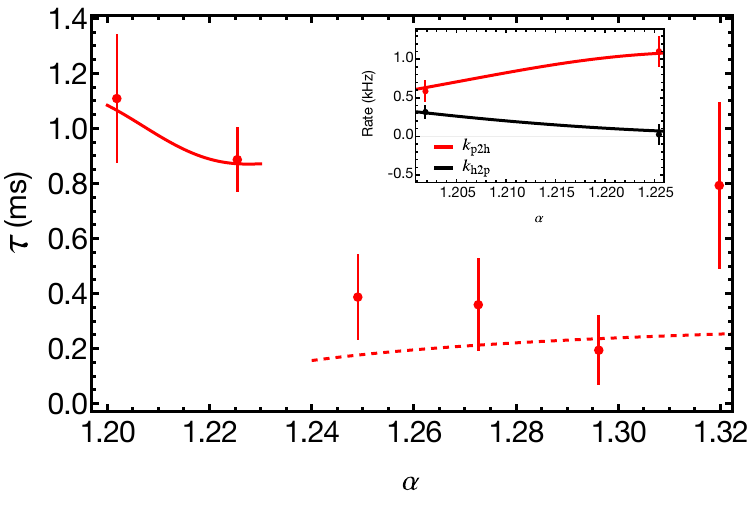}
    \caption{\textbf{Isomerization time as a function of the trap aspect ratio.} The red points report the isomerization characteristic time $\tau$ as a function of $\alpha$. The times are obtained from an exponential fit of the photon rate data (see Fig. \ref{fig:decay}c). From the simulation, a double well is predicted for $\alpha$ below approx. $1.23$. The solid red line corresponds to the decay expected from a double well model of isomerization. The dashed line corresponds to the time needed for recooling the crystal below the energy on the potential energy surface corresponding to the edge of the detection region (see text). Inset: transition rates $k_{\mathrm{h2p}}$ (black) and $k_{\mathrm{p2h}}$ (red). The corresponding solid lines represent the expected values for the PES extracted in the Monte Carlo simulation.
   }
    \label{fig_tau}
\end{figure}

We use Eq.s (\ref{Eq:probfitting}) to fit the experimental data in Fig. \ref{fig_probability}. To this end, we extrapolate the data of $V_{\mathrm{h2p}}$ and $V_{\mathrm{p2h}}$ from the Monte Carlo simulation, and use the temperature as the only free parameter. We change the temperature in steps of $0.1$ \si{\milli\kelvin}, and identify the curve that has the lowest $\chi^2$, finding as the best temperature $8.8$ \si{\milli\kelvin} for both $p_{\mathrm{hex}}$ and $p_{\mathrm{pen}}$ (see solid lines in Fig. \ref{fig_probability}). To estimate the uncertainty of this temperature estimate, we consider the temperatures for which the $\chi^2$ parameter of the corresponding function is a factor of $2$ larger than the best fitting value. The corresponding temperature range is $8.8 \substack{+3.5 \\ -2.6}$ \si{\milli\kelvin}. This range is shown in Fig. \ref{fig_probability} as a shaded area.
We further investigate isomerization by analyzing the transient behaviour. By analyzing the data after the aspect ratio quenching, we extract the time constant $\tau$ of the decay of the photon rate (see Fig. \ref{fig:decay}c).
Fig. \ref{fig_tau} shows the scaling of $\tau$ as a function of the trap aspect ratio. We observe a faster decay as the aspect ratio is increased.
From the Monte Carlo simulation, we can identify two separate regions: 
for $\alpha \lesssim 1.23$ the system is bistable, as we observe two minima in the corresponding PES, while for $\alpha \gtrsim 1.23$ only the hexagon isomer is stable.
In the bistable region, the decay time is related to the rates of the transitions between the two energy minima by the formula  $\tau=\left( k_{\mathrm{p2h}} + k_{\mathrm{h2p}}\right)^{-1}$\cite{RevModPhys.62.251}. By using Eq. (\ref{eq:prob}), we can extract the relation of the rates: 
\begin{equation}
\begin{aligned}
k_{\mathrm{p2h}}&=\left( \tau (1 + p_{\mathrm{pen}}/p_{\mathrm{hex}}) \right)^{-1},
\\
k_{\mathrm{h2p}}&=\left( \tau (1 + p_{\mathrm{hex}}/p_{\mathrm{pen}}) \right)^{-1}.
\end{aligned}
\label{Eq:ratesformula}
\end{equation}
We use Eq.s (\ref{Eq:ratesformula}) and the best fitting curves of the data in Fig. \ref{fig_probability} to extract the rates from the observed time constant. The results are shown in the inset of Fig. \ref{fig_tau}, where the solid lines are a fit of the data with the Arrhenius equation. The best fitting functions of the rates are used to extract the expected behaviour for the time constant $\tau$, shown in Fig. \ref{fig_tau} as a solid red line. We find an excellent agreement with the experimental data.
Out of the bistability region, i.e. for $\alpha \gtrsim 1.23$, we observe a non-negligible decay time that is long compared to the eigen-frequencies of the vibrational modes of the isomers.
We attribute this time to be caused by the relatively high kinetic energy acquired by the ions after the quenching, which increases the time required to cool the crystal.
From the PES, we can calculate the excess energy the crystal has to lose so that no ion passes by the region of space observed by our imaging system. The dashed line in Fig. \ref{fig_tau} shows the time needed by a free particle to lose the excess energy for the cooling light parameters used in the experiment. The curve qualitatively confirms our interpretation.

\subsection*{Conclusions}
In this work, we demonstrate that Coulomb crystals can be used to emulate the physics of isomerization, and study the bistabilities that may arise.
Thanks to the high level of control on the experimental parameters, we have selectively arranged the crystal in a metastable isomer, thus emulating the physics of isomerization of a molecule.
By exploiting the slow dynamics when compared to real molecules and a configuration-resolved detection, we have identified the bistability region and revealed transitions between two isomers that are excited thermally.
The potential energy surface of the crystal is obtained using a Monte Carlo simulation and used to derive the barrier heights and the  transition rates.
Data obtained from the simulation shows an excellent agreement with the experimental data and are used to evaluate the temperature of the crystal.
Our study paves the way towards exploiting trapped ions for simulations of fundamental mechanism in chemical reactions, e.g. by further modifying the crystal PES by using optical potentials \cite{horak2012} or more complex electric potentials \cite{Mielenz2016}.
Moreover, understanding the hysteresis of the isomer formation may play an essential role in the initial state preparation of a quantum computer or a quantum simulator realized with two-dimensional Coulomb crystals \cite{Kiesenhofer2023, Guo2024}.
Finally, our ability to control the relative height of the wells in the observed double-well potential demonstrates that this system can serve as a new platform for realizing quantum superpositions of crystal isomers \cite{retzker2008} and observing quantum tunneling between them \cite{Noguchi2014}. 

While preparing this manuscript, we became aware of one study \cite{ayyadevara2025} reporting transitions between two stable configurations of trapped ions in three dimensions, and a second study \cite{mallweger2025} discussing trapped ions in quasi-one-dimensional confinement in a similar context.

\section*{Methods}

\subsection*{Monte Carlo simulation}
We have conducted Monte Carlo simulations to identify the metastable configuration of the ions and their corresponding potential energy surface.
The ions are first placed in a two-dimensional harmonic potential of frequencies ($\omega_z$, $\omega_y$) in a 1 \si{\nano\metre} grid. Then, the ions positions are optimized by moving two ions at a time within a $10$ \si{\nano \metre} range in the plane.
We update the positions of the two ions to the ones that minimizes the energy of the whole system.
The procedure is repeated until we get no improvement in the energy after $150$ random trials.
For each set of parameters, a simulation was repeated for at least three times.
The trap frequencies are varied according to the experimentally determined dependence of the aspect ratio on the DC voltage which was measured with a parametric heating experiment.
To compensate for the small errors e.g. due to surface charges on the trap electrodes, we calibrated the trap frequencies by comparing the experimental data and the simulation in the condition in which barrier heights between two isomers are balanced.
For the Monte Carlo simulation in Fig. \ref{fig:bistability}a, the simulation becomes less accurate for the metastable isomer around the edges of the bistable region.
This is because the PES becomes very flat, and even if there is no minimum in the PES the simulation terminates before converging to a stable solution.
To avoid such issues, we assessed the stability of the ion equilibrium positions obtained from Monte Carlo simulations by analyzing the eigenvalues of the Hessian matrix.
The presence of a negative eigenvalue indicates that the configuration does not correspond to a true equilibrium.

\subsection*{Preparation and detection of two-dimensional ion crystals}
To confine ions in a two-dimensional plane, our Paul trap has a specific design in which DC electrodes are negatively charged \cite{duca2023}.
By changing the voltages applied to the DC electrodes, the confining potential can vary from 1D to 2D.
For example in Fig. \ref{fig:decay}, confining potential is quenched from $(\omega_z,\omega_y)/(2\pi) = (98.777,127.849)$ \si{\kilo \hertz} 
to $(\omega_z,\omega_y)/(2\pi) = (99.691,122.144)$ \si{\kilo \hertz}.
Ions are laser cooled with two laser beams at 493.5 nm and a repumping laser of 650 nm corresponding to the $\ket{6s \, ^{2}\mathrm{S}_{1/2}} \rightarrow \ket{6p \, ^{2}\mathrm{P}_{1/2}}$ and $\ket{5d \, ^{2}\mathrm{D}_{5/2}} \rightarrow \ket{6p \, ^{2}\mathrm{P}_{1/2}}$ transitions respectively.

We detect fluorescence at 493.5 nm 
from the two opposite sides on the axis orthogonal to the trapping plane.
On one of the sides, we use two telescopes, each with a magnification of 2, and use a polarizing beam splitter for splitting the signal and send it to a photon counter and a charge-coupled device (CCD) camera.
In the first image plane after the first telescope, a pinhole (20 \si{\micro \metre} diameter) is placed to detect the photons coming from the center region (10 \si{\micro\metre} diameter) of the trap.
The pinhole was aligned by trapping a  single ion acting as a reference for the trap center, and looking at the picture in the CCD camera.
The photon arrival time is recorded with a time-to-digit converter and counted with bins of 0.1 \si{\milli \second}.
We monitor the number of ions in the trap with another CCD camera from the opposite side of the plane.

\begin{acknowledgments}
We thank Stefan Willitsch and Massimo Inguscio for helpful discussions. This work has been supported by the project 23FUN03 HIOC, which has received funding from the European Partnership on Metrology, cofinanced from the European Union’s Horizon Europe Research and Innovation Programme and by the Participating States. This work was supported by the Cascading Grant Spoke 3 under the PNRR MUR project PE0000023-NQSTI, funded by the European Union – Next Generation EU.
\end{acknowledgments}

\appendix

\bibliography{references}

\end{document}